\documentclass{ws-procs975x65}
\def\ltsima{$\; \buildrel < \over \sim \;$}
\def\simlt{\lower.5ex\hbox{\ltsima}}            % < over ~
\def\gtsima{$\; \buildrel > \over \sim \;$}
\def\simgt{\lower.5ex\hbox{\gtsima}}            % > over ~

\def\grcm3{g\,cm$^{-3}$}
\def\gra{^{\circ}}

\begin{document}

%to switch ON running title
%\markboth{L. Hatcher}{Quantum States from Tangent Vectors}

%\wstoc{Quantum States from Tangent Vectors}{L. Hatcher}

\title{MAKING UP A SHORT GRB: THE BRIGHT FATE OF MERGERS OF COMPACT
  OBJECTS\footnote{This research has been partially supported by the
    Spanish Ministerio de Ciencia y Tecnolog\'{\i}a grant
    AYA2001-3490-C02-C01 and through the EU grant
    MEIF-CT-2005-009395.}}

\author{MIGUEL-ANGEL ALOY\footnote{Ram\'on y Cajal Fellow of the
    Spanish Ministry of Education and Science} and
  P. MIMICA\footnote{Marie Curie Fellow}}

\address{Departamento de Astronom\'\i{}a y Astrof\'\i{}sica,
  Universidad de Valencia \\
 46100, Burjassot, Spain \\
%and\\
%Max-Planck-Institut f\"ur Astrophysik,\\
% Karl-Schwarschild-Str. 1, 85741 Garching, Germany\\
\email{Miguel.A.Aloy@uv.es}}

% WARNING. in standard latex cls file formatting, at this point
% \maketitle would typeset the above titlepage information
% but WS has chosen to be nonstandard and have each line typeset 
% as it is digested.
% no abstract is necessary.
% \bodymatter below resets the footnote counter and symbols after 
% possible use in the title matter.

\begin{abstract}
  We show some of the most important reasons why the likely fate of
  the merger of a neutron star with another compact object may be to
  yield a short gamma-ray burst (sGRB).  Emphasis is made on some
  robust results that general relativistic (magneto)hydrodynamic
  simulations have established regarding the aforementioned subject.
\end{abstract}

\bodymatter

\section{What do we know about the progenitor systems of sGRBs?}
\label{sec:intro}

Our knowledge about the progenitors of short gamma-ray bursts is
relatively small. Any proposed progenitor system should be able of
releasing $\sim 10^{49}\,$erg in the form of thermal energy or
Poynting flux. Furthermore, the progenitor is requested to yield
outflows collimated into cones of half-opening angle
$\theta_\mathrm{j} \sim 4\gra - 25\gra$, as inferred from the observed
breaks in the light curves of some short GRBs,\cite{Burrowsetal06} or
the lower limits on $\theta_\mathrm{j}$ deduced from the absence of
such breaks.\cite{DeUgarteetal06,Grupeetal06}. Additionally, flow
variability down to a few milliseconds over timescales $\sim 0.1 -
2\,$s has to be produced. Also, any viable progenitor has to satisfy
the fact that it can be generated in any of the {\em typical} hosts
observed for a handful of burst\cite{Prochaskaetal06}, as well as to
occur at rates equal or larger than the rate at which sGRBs are
produced.

\section{Why are mergers of compact objects good candidates to be
  progenitors of  sGRBs?}

A family of systems that has been proposed as likely progenitors of
sGRBs is the remnant left by the merger of a binary system of compact
objects formed by either two neutron stars (NSs) or a NS and a
BH.\cite{merger_grb} Such systems consist of a BH girded by a thick
gas torus from which it swallows matter at a hypercritical rate. In
this situation the cooling is dominated by the emission of
neutrinos. These neutrinos might either be the primary energy source
blowing a fireball of e$^+$e$^-$ pairs and photons or to load with
pairs a Poynting dominated outflow.\cite{LE93} The duration of the
produced outflows is, in part, limited by the time needed by the BH to
engulf most of the matter of the accretion disk, namely, a few 100\,ms
(but see below).  Thereby, a merger of compact objects (MCO) may
release energy during a sufficiently long time to fuel a {\it short}
duration sGRB event.  This limit on the time scale, set by the ON time
of the source, holds for both $\nu$-powered jets and for MHD-generated
outflows.\cite{AO07}

According to the state-of-the-art numerical simulations of MCOs
including realistic microphysics\cite{merger_sim} it is possible to
release a few $10^{49}\,$erg above the poles of a stellar mass BH in a
region of nearly vacuum as a result of the process of
$\nu\bar\nu$-annihilation in such region. Detailled simulations of
such process suggest that even more energy could be deposited in the
system\cite{BAJM06}, although time-dependent numerical models
including energy transport are needed to give more reliable
numbers. Therefore, MCOs provide an energy budged which may satisfy
the energy requirements to produce sGRBs. Numerical simulations have
also shown the ability of the system to tap the energy from the BH and
potentially fuel a GRB.\cite{McKinney06}

In the standard merger scenario, initially the spiral-in process may
take $\sim 1\,$Gy. Since it is likely that the newly born binary
system may receive a natal kick (of a few $\sim 100\,$km\,s$^{-1}$),
it is expected that, when the two compact objects merge, they have
travelled to the outer skirts of the galaxy or, even, to the
intergalactic medium. This evolutionary path is compatible with
observing sGRBs outside of their putative host galaxies (e.g.,
GRB~050813)\cite{Prochaskaetal06}. Alternatively, the evolutionary
tracks of MCOs might be much shorter ($\sim 1\,$My)\cite{fastmergers},
which has the implication that many mergers could happen inside of
their hosts galaxies and, hence, this would accommodate the small
offsets with respect to the galactic center observed for a few sGRBs
(e.g., GRB~050724)\cite{Prochaskaetal06}.

The number of well identified galaxies hosting sGRBs is still small to
have a good statistical sample. Nevertheless, it seems that sGRBs are
associated with both (young) start-forming and (old) elliptical
galaxies.\cite{Prochaskaetal06,Foxetal05} It is very appealing (but
still quite speculative) to establish an association between sGRBs
detected in old galaxies with mergers happening after long
evolutionary paths and, on the other side, between sGRBs found in
young galaxies with faster evolving mergers. However, taking together
the information provided by the typical galactic offsets and the
typical host galaxies of sGRBs one may infer only circumstantial
evidences about the nature of the progenitors of sGRBs.

Self-consistent (magneto)hydrodynamic modeling is needed to address
issues like:

\noindent 
{\bf Collimation}.  The generated outflows are either collimated by
the accretion disk\cite{AJM05} or self-collimated by the magnetic
field \cite{McKinney06}. The typical outflow half-opening angles are
$\sim 3^o - 25^o$ (i.e., compatible with observations; see
\S~\ref{sec:intro}). The baryon-poor outflows develop a transverse
structure.  Particularly, the transverse profile of the Lorentz factor
could be roughly fit by a Gaussian function, but more complicated
functions are required to provide accurate fits.\cite{AJM05}

\noindent
{\bf Variability}. Even injecting energy at constant rates, the
produced outflows are highly variable. The interaction of the newborn
fireball with the accretion torus yields the growth of
Kelvin-Helmholtz \cite{AJM05} instabilities.  In case of MHD jets the
variability is imprinted by pinch instabilities.\cite{McKinney06} All
computed axisymmetric models seem to be either stable or marginally
stable. It is not yet numerically verified whether 3D jets emerging
from hyperaccreting BHs are stable.

\noindent
{\bf Influence of the environment}. Mergers in low density
environments may fuel ultrarelativistic outflows with the potential to
produce {\it normal} sGRBs, while in case that the merger occurs in
high density media, the observational signature is not a
sGRB.\cite{AJM05} The fact that depending on the environmental density
an sGRB can be produced or not has direct implications for the
estimates\cite{GP05} of the true rates of sGRBs and compared with the
rates of NS+NS mergers.

\noindent
{\bf Asymptotic Lorentz factor}. Saturation values of the bulk Lorentz
factor, $\Gamma_\infty\simgt 500 - 1000$ are obtained for both
thermally\cite{AJM05} and magnetically\cite{McKinney06} generated
outflows. For thermally produced outflows, there is a trend to produce
much higher values of $\Gamma_\infty$ for sGRBs ($\Gamma_\infty \simgt
500 - 1000$) than for lGRBs ($\Gamma_\infty \sim 100$). This
difference in Lorentz factor might be the reason for the paucity of
{\em soft} sGRBs.\cite{Jetal06}

\noindent
{\bf Duration of the events}.  In addition to the ON time of the
source, the other factor that sets the duration of a GRB event is the
radial stretching of the fireball which results from the differential
acceleration of the forward and rear edges of the
fireball.\cite{AJM05,Jetal06}. A prolonged activity could also be
produced by the fall back onto the BH of a fraction of the matter
ejected during the early stages of the merger.\cite{Rosswog06}

\section*{Acknowledgments}
It is a pleasure to thank H.-Th Janka and E. M\"uller for
encouragement and support.  The author thanks Max-Planck-Institut
f\"ur Astrophysik for hospitality.  This work has been partially
supported\- by SFB 375 ``Astroparticle Physics''and SFB-TR 7
``Gravitational Wave Astronomy'' of the Deutsche
Forschungsgemeinschaft.
% \eject

\vfill
%\pagebreak

\end{document}